\documentclass[prd,onecolumn]{revtex4}
\usepackage{dcolumn}
\usepackage{multirow}
\usepackage{graphicx}
\usepackage{amssymb}
\usepackage{bm}
\usepackage{hyperref}
\usepackage{epstopdf}
\usepackage{color}
\usepackage{mathrsfs}
\usepackage{amsmath,amssymb,amsthm}
\usepackage{rotating}
\usepackage{sverb, longtable}
\usepackage{subfigure}

\usepackage[graphicx]{realboxes}
\usepackage{adjustbox}
\begin{document}

\title{Dynamical dark energy can explain the anomalous matter density fluctuation at $z\sim4$}
\author{Deng Wang}
\email{cstar@nao.cas.cn}
\affiliation{National Astronomical Observatories, Chinese Academy of Sciences, Beijing, 100012, China}
\begin{abstract}
Recently, under the standard cosmological model, a new growth tension between the Planck-2018 observation and the combined observation of cosmic microwave background lensing and galaxy clustering at $z\sim4$ emerges over the $1\,\sigma$ confidence level. We demonstrate that dynamical dark energy can well solve this tension within the $1\,\sigma$ confidence level. This implies that the new measurement of large scale structure at high redshift may give the evidence of evolution of dark energy over time.

\end{abstract}
\maketitle

\section{Introduction}
During the past two decades, the standard $\Lambda$-cold dark matter ($\Lambda$CDM) model has been successfully confirmed by a large number of observations such as Type Ia supernovae (SNe Ia) \cite{SupernovaSearchTeam:1998fmf,SupernovaCosmologyProject:1998vns}, baryon acoustic oscillations (BAO) \cite{Blake:2003rh,Seo:2003} and cosmic microwave background (CMB) \cite{WMAP:2003elm,Planck:2013pxb,Planck:2018vyg}. However, it is imperfect and faces at least four intractable problems: (i)
the locally direct measurement of the Hubble Constant $H_0$ from the Hubble Space Telescope (HST) \cite{Riess:2021jrx} is $5\,\sigma$ higher than the globally derived $H_0$ value from the Planck-2018 CMB data under the assumption of $\Lambda$CDM \cite{Planck:2018vyg}; (ii) present-day matter density fluctuation amplitude $\sigma_8$ measured by several low-redshift probes including weak gravitational lensing \cite{Aghanim:2016yuo}, cluster counts \cite{Battye:2014qga} and redshift space distortions \cite{Macaulay:2013swa} is lower than that indirectly measured by the Planck-2018 CMB observations under $\Lambda$CDM \cite{Planck:2018vyg}; (iii) the value of cosmological constant from current observations is much smaller than that from the theoretical calculation; (iv) why the present-day densities of dark matter and dark energy (DE) are of the same order \cite{Weinberg:1988cp}? These pendent problems have initiated intense discussions of possible new physics and underlying systematic uncertainties of the measurements in the community (see Refs.\cite{Abdalla:2022yfr,DiValentino:2020vvd,DiValentino:2020zio} for recent reviews).  

Recently, Hironao {\it et al.} \cite{Miyatake:2021qjr} (hereafter H22) report the first high-redshift (high-z) measurement of the matter density fluctuation at $z\sim4$ by combining the CMB lensing with galaxy-galaxy clustering observations, the latter of which is obtained from 1473106 Lyman break galaxies with the median redshift of $z\sim4$. This high-z galaxy sample covering a sky area of 305 deg$^2$ is observed by the Hyper Superime-Cam Strategic Survey Program survey \cite{Aihara:2017paw}. Interestingly, they find that the constraints on $\sigma_8$ and $f\sigma_8$ ($f$ is the linear growth rate) under $\Lambda$CDM at $z\sim4$ are lower than that from the Planck-2018 cosmology \cite{Planck:2018vyg} over the $1\,\sigma$ confidence level. This implies that there is a small tension between these two probes in the framework of $\Lambda$CDM. In this study, we will show that this small growth tension can be well solved by dynamical dark energy (DDE) models (see Refs.\cite{Wang:2022xdw,Wang:2019ufm,Wang:2017jdm,Xu:2019amr,Zhao:2017cud,Escamilla-Rivera:2021boq} for more details). As a consequence, this new measurement of large scale structure at high redshift may give the evidence of evolution of DE over time.

This study is structured as follows. In the next section, we review briefly the specific formula of the considered DDE model. In Section III, we introduce the analysis methodology and display the numerical results. Discussions and conclusions are presented in the final section. We use the units $8\pi G=c=1$ throughout this study.    

\section{DDE model}
In the framework of general relativity, a flat, homogeneous and isotropic universe is described by the Friedmann-Robertson-Walker (FRW) metric
\begin{equation}
\mathrm{d}s^2=-\mathrm{d}t^2+a^2(t)\left(\mathrm{d}r^2+r^2\mathrm{d}\theta^2+r^2\mathrm{sin}^2\theta \mathrm{d}\phi^2\right),      \label{1}
\end{equation}
where $a(t)$ is the scale factor at cosmic time $t$. Inserting Eq.(\ref{1}) into the Einstein field equation, we obtain the Friedmann background dynamics equations   
\begin{equation}
H^2=\frac{\sum\rho_i}{3},     \label{2}
\end{equation}   
\begin{equation}
\frac{\ddot{a}}{a}=-\frac{\sum(\rho_i+3p_i)}{6},     \label{3}
\end{equation}   

where $H$ is the Hubble parameter, and $\rho_i$ and $p_i$ represent the mean energy density and mean pressure of each component $i$. Specifically, in this study, we consider the matter and dark energy components in the cosmic pie since just focusing on the late-time universe. Subsequently, combining Eq.(\ref{2}) with Eq.(\ref{3}), we show the energy conservation equation as 
\begin{equation}
\dot{\rho_i}+3\frac{\dot{a}}{a}(\rho_i+p_i)=0.     \label{4}
\end{equation} 
This equation can also be easily derived from $\Delta_\mu T^{\mu\nu}=0$. Assuming $p_i=\omega_i\rho_i$, where $\omega_i$ denotes an equation of state (EoS) for each component, one can easily derive the dimensionless Hubble parameter (DHP) $E(z)$, which describes the background dynamics of a cosmological model, for the $\Lambda$CDM model
\begin{equation}
E_{\mathrm{\Lambda CDM}}(z)=\left[\Omega_{m}(1+z)^3+1-\Omega_{m}\right]^{\frac{1}{2}}, \label{5}
\end{equation}
where $z$ and $\Omega_m$ and $\Omega_K$ are the redshift, present-day matter fraction, respectively.

To study the dynamics of DE, the simplest model should be $\omega$CDM, a one-parameter extension to $\Lambda$CDM, where DE is characterized by a barotropic fluid with a constant EoS $\omega(z)=\omega$. The DHP of the flat $\omega$CDM model reads as
\begin{equation}
E_{\mathrm{\omega CDM}}(z)=\left[\Omega_{m}(1+z)^3+(1-\Omega_{m})(1+z)^{3(1+\omega)}\right]^{\frac{1}{2}}. \label{6}
\end{equation}
It is clear that $\omega$CDM reduces to $\Lambda$CDM when $\omega=-1$. 

Furthermore, to explore whether DE evolves over time, a simple approach is considering the DDE scenario, which assumes the DE EoS as a function of redshift. Specifically, we consider the so-called Chevallier-Polarski-Linder model $\omega(a)=w+w_a(1-a)$ \cite{Chevallier:2000qy,Linder:2002et}, where the parameter $w_a$ describes the evolution of DE. This model is aimed at solving the problematic behavior at high redshifts, gives an excellent fit for
a number of theoretically conceivable scalar field potentials, and provide a good explanation for a small deviation from the
phantom barrier $\omega=-1$. Simultaneously, $\omega(z)$ is well behaved at $z\gg 1$ and recovers the
linear behavior at low redshifts. Hereafter we call this model as DDE. The DHP of the flat DDE model is expressed as 
\begin{equation}
E_{\mathrm{DDE}}(z)=\left[\Omega_{m}(1+z)^3+(1-\Omega_{m})(1+z)^{3(1+\omega+\omega_a)\mathrm{e}^{\frac{-3\omega_az}{1+z}}}\right]^{\frac{1}{2}}, \label{7}
\end{equation}
which reduces to $\Lambda$CDM when $\omega=-1$ and $\omega_a=0$ .

Considering the scalar perturbations only, using the conformal Newtonian gauge, the perturbed FRW spacetime reads as \cite{Mukhanov:1992,Ma:1995,Malik:2009}
\begin{equation}
ds^2=a^2(\tau)\left[ -(1+2\Psi)d\tau^2+(1-2\Phi)\gamma_{ij}dx^idx^j  \right], \label{8}
\end{equation}
where $\Psi$ and $\Phi$ are perturbed metric potentials and $\tau$ denotes the conformal time. The components of perturbed energy-momentum tensor are expressed as 
\begin{equation}
\delta T_0^0=-\tilde{\delta}\rho, \label{9}
\end{equation}
\begin{equation}
\delta T_0^i=-(1+c_s^2)\rho v^i, \label{10}
\end{equation}
\begin{equation}
\delta T_1^1=\delta T_2^2=\delta T_2^2=c_s^2\tilde{\delta}\rho, \label{11}
\end{equation}
where $\rho$ is the density of the fluid, $\tilde{\delta}=\delta\rho/\rho$ is the dimensionless density perturbation, $v$ is the velocity perturbation, and $c_s$ is the adiabatic sound speed of the fluid. 

For the non-relativistic matter component, its EoS $\omega_m$ and squared adiabatic sound speed $c^2_{s(m)}$ both equal zero, namely $\omega_m=c^2_{s(m)}=\delta P_m/\delta \rho_m=0$. Subsequently, for the matter component, the temporal and spatial components of perturbed gravitational field equation are expressed as 
\begin{equation}
\delta_m^\ast=3\Phi^{\ast}-\theta_m, \label{12}
\end{equation} 
\begin{equation}
\theta_m^\ast=k^2\Psi-\mathcal{H}\theta_m, \label{13}
\end{equation} 
where $\mathcal{H}$ is the conformal Hubble parameter, the symbol ``$\ast$'' denotes the derivative with respect to the conformal time, and $\delta_m$ and $\theta_m$ denote the density and velocity perturbations of matter, respectively. 

Inserting the relation $\Psi=\Phi$ derived from the spatial off-diagonal component of perturbed field equation into Eqs.(\ref{12}) and (\ref{13}), in the sub-horizon limit $k\gg \mathcal{H}$ ($k$ is the comoving wave number), one can easily obtain a second order differential equation for the matter density perturbation 
\begin{equation}
\delta_m^{\prime\prime}+(\frac{H^\prime}{H}-\frac{1}{1+z})\delta_m^\prime-
\frac{3\Omega_{m0}(1+z) }{2E^2(z)}\delta_m = 0 , \label{14}
\end{equation}
where the prime is the derivative with respect to the redshift $z$. 

In order to study the behaviors of large scale structure in the DDE model, we introduce the following perturbation quantity
\begin{equation}
f\sigma_8(z)=\sigma_8(z)\frac{\delta_m^\prime}{\delta_m} ,\label{15}
\end{equation}
where $\sigma_8(z)\equiv\sigma_8\delta_m(z)/\delta_m(0)$. In the next section, we will calculate the redshift-dependent $f\sigma_8(z)$ for the DDE model at $z\sim4$ based on the Planck-2018 constraint.

\begin{figure}
	\centering
	\includegraphics[scale=0.6]{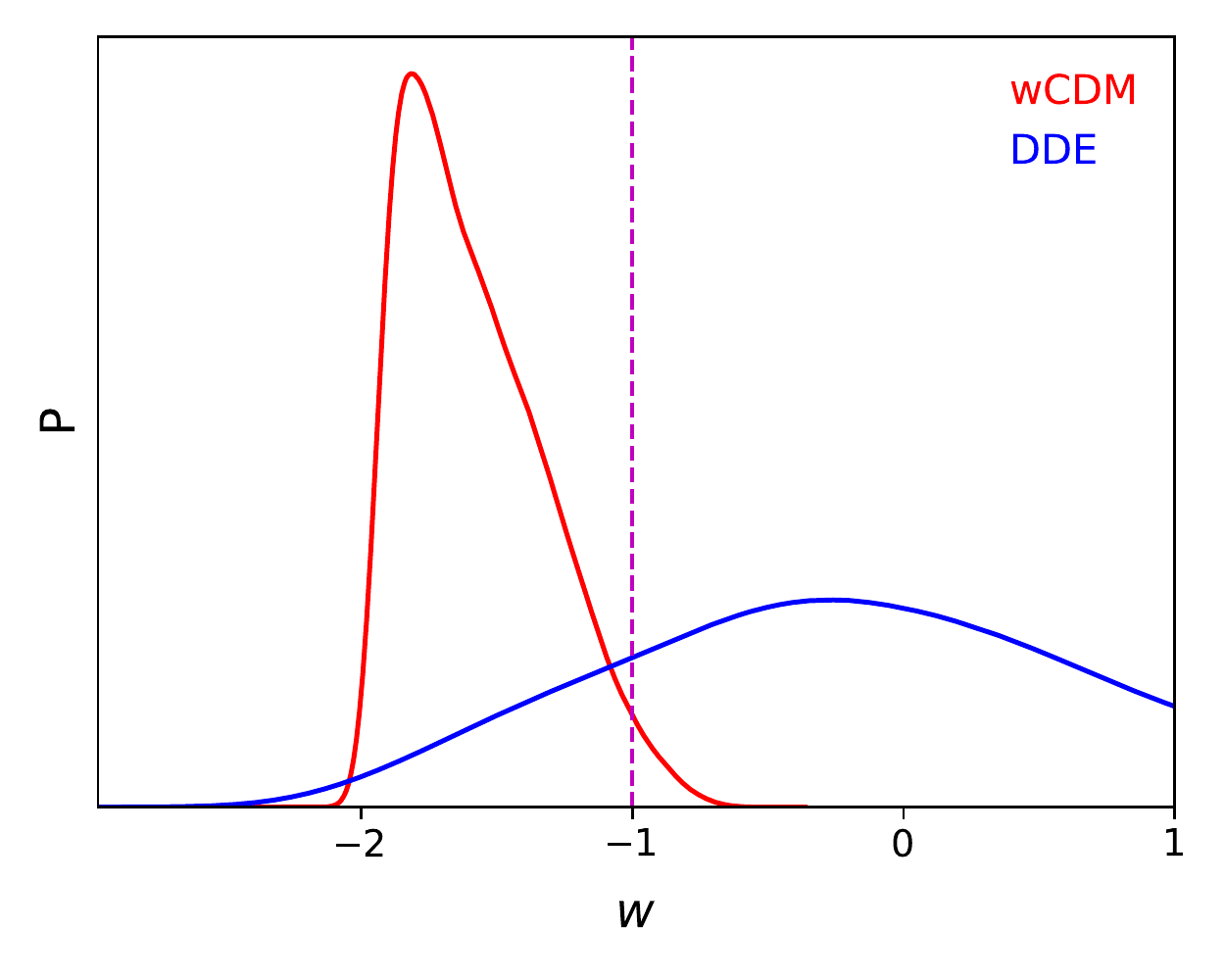}
	\includegraphics[scale=0.6]{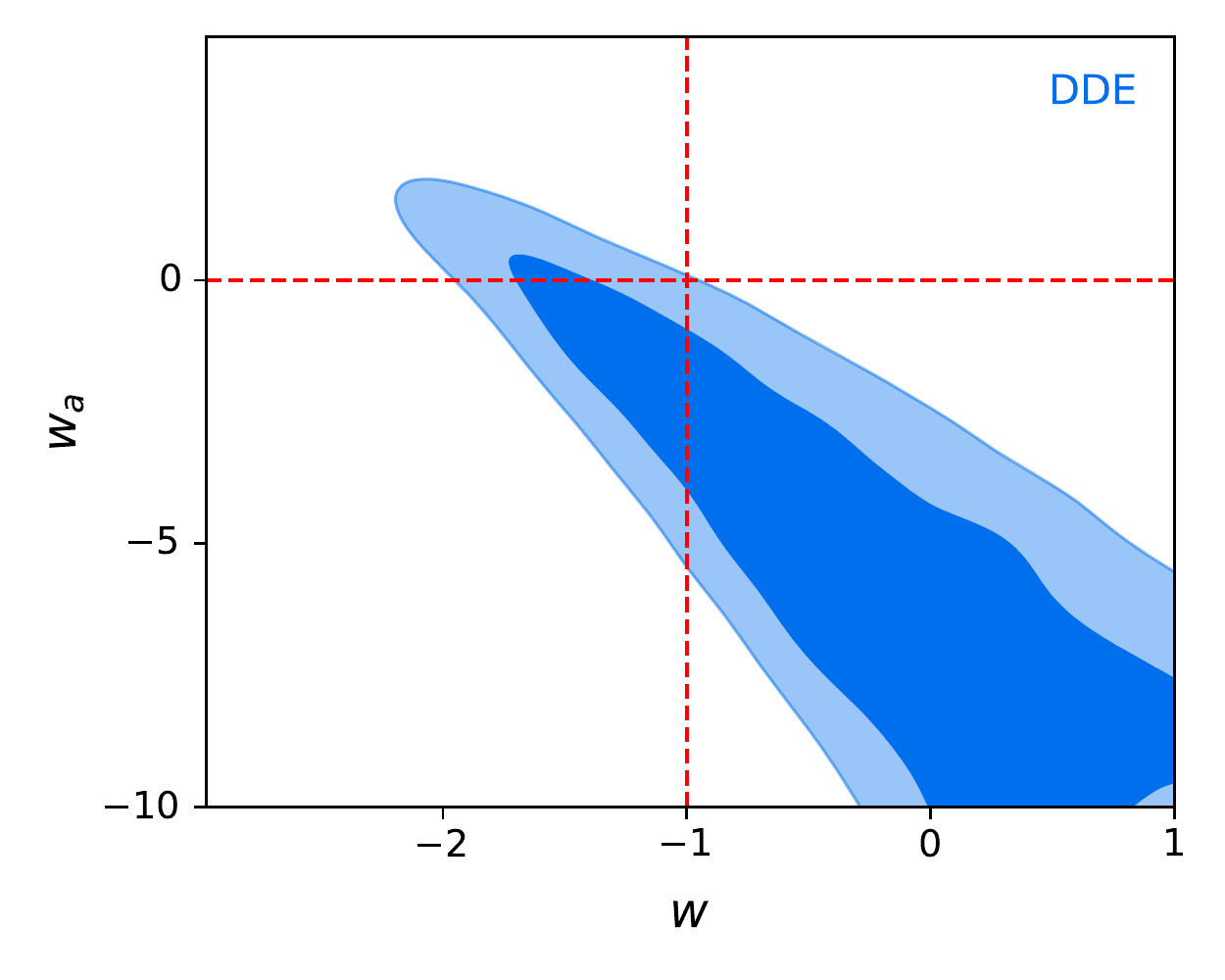}
	\caption{{\it Left}. The marginalized 1-dimensional posterior distribution of DE EoS $\omega$ in the $\omega$CDM and DDE models. The magenta dashed line corresponds to $\Lambda$CDM. {\it Right}. The 2-dimensional posterior parameter space ($\omega,\,\omega_a$) of the DDE model. The cross point of two red dashed lines denotes $\Lambda$CDM.}\label{f1}
\end{figure}

\begin{figure}
	\centering
	\includegraphics[scale=0.7]{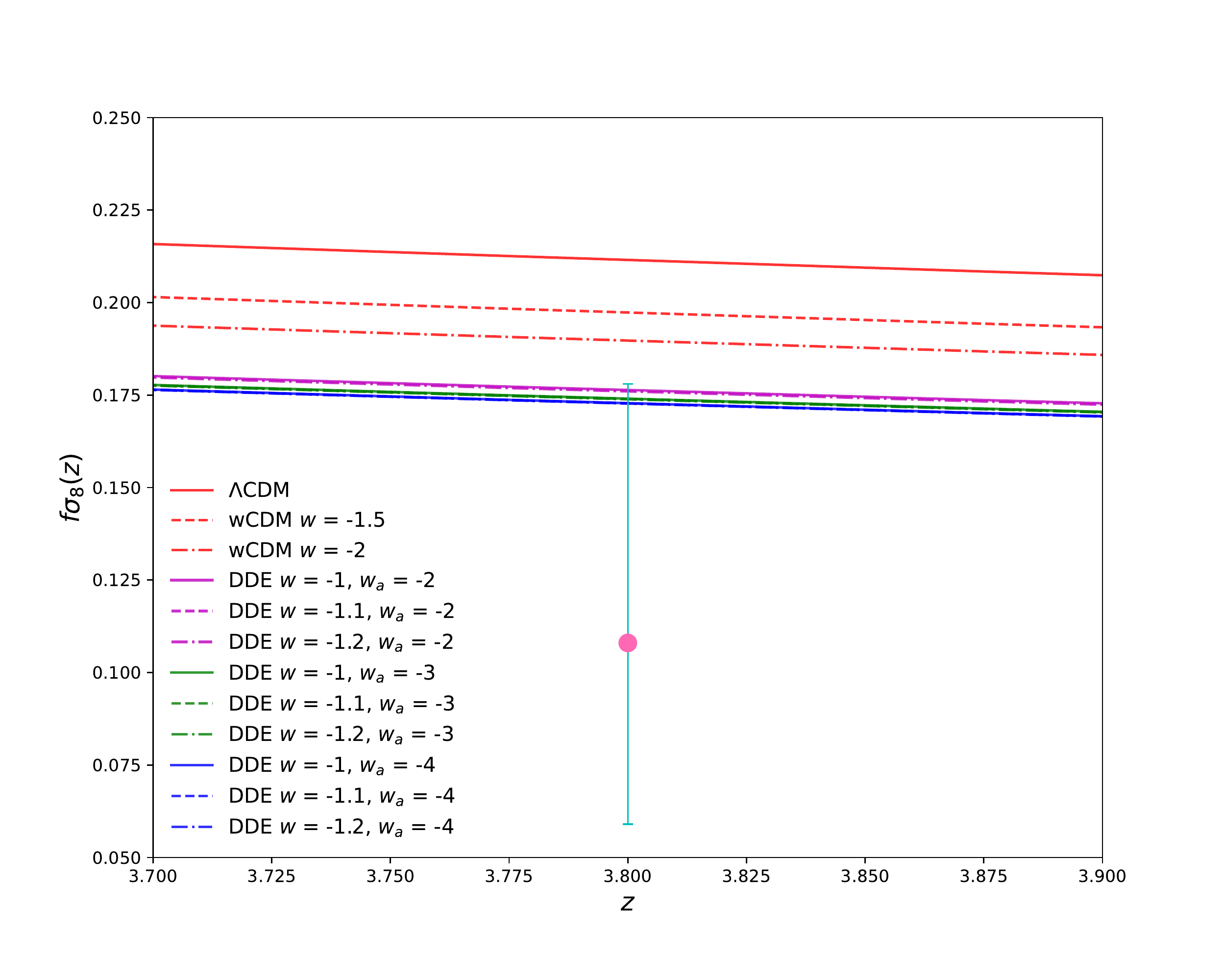}
	\caption{$f\sigma_8(z)$ as a function of redshift $z$ for $\Lambda$CDM, $\omega$CDM and DDE models around $z=3.8$. Different lines denote different model predictions permitted by Planck-2018 observations within $1\,\sigma$ confidence level. The data point with error bar is the measurement from Ref.\cite{Miyatake:2021qjr} at $z=3.8$.}\label{f2}
\end{figure}

\section{Methodology and results}
The analysis methodology is simply divided into the following two steps: (i) give the allowed parameter space of DE EoS by constraining $\omega$CDM and DDE models with the Planck-2018 data; (ii) use the allowed parameter values of DE EoS to calculate $f\sigma_8(z)$ for $\omega$CDM and DDE models at $z\sim4$.

The Planck-2018 observation have very important implications for cosmology. It can help measure the matter components, topology and large scale structure of the universe. To place constraints on the above two models, we use the Planck-2018 CMB temperature and polarization data including the likelihoods of temperature at $30\leqslant \ell\leqslant 2500$ and the low-$\ell$ temperature and polarization likelihoods at $2\leqslant \ell\leqslant 29$, namely TTTEEE$+$lowE \cite{Planck:2018vyg}, and Planck-2018 CMB lensing data.

In order to implement the constraints, we employ the public package \texttt{CosmoMC} \cite{Lewis:2013hha} which takes a standard Bayesian analysis via the Markov Chain Monte Carlo (MCMC) sampling method to obtain the posterior distributions of free parameters. Then, we use \texttt{Getdist} \cite{Lewis:2019xzd} to analyze the MCMC chains. The Gelman-Rubin statistic $R-1=0.01$ is taken as the convergence criterion of our MCMC analysis. For $\omega$CDM, we use the priors $\Omega_bh^2 \in [0.005, 0.1]$, $\Omega_ch^2 \in [0.001, 0.99]$, $100\theta_{MC} \in [0.5, 10]$,  $\mathrm{ln}(10^{10}A_s) \in [2, 4]$, $n_s \in [0.8, 1.2]$, $\tau \in [0.01, 0.8]$, $\omega \in [-3, 1]$, where $\Omega_bh^2$ and $\Omega_ch^2$
are the present-day baryon and CDM fraction, $\theta_{MC}$ is the ratio between angular diameter
distance and sound horizon at the last scattering redshift, $\tau$ is the optical depth due to the reionization, and $A_s$ and $n_s$ are the amplitude and spectral index of primordial scalar power spectrum. For DDE, we add the prior $\omega_a \in [-10, 5]$ into the parameter priors of $\omega$CDM.

The constraining results of $\omega$CDM and DDE models are shown in Figs.\ref{f1}. One can easily find that Planck-2018 observation prefers a phantom energy in the case of $\omega$CDM. However, when we consider the evolution of DE over time, this preference disappears in the case of DDE. It is interesting that the allowed $\omega$ range is enlarged for DDE and the parameter space $(\omega,\,\omega_a)$ is large. As a consequence, we expect that $f\sigma_8(z=3.8)$ derived from Planck can be consistent with that derived from the H22 measurement in the DDE model. 

Our numerical results are presented in Fig.\ref{f2} by adopting different values of model parameters of $\omega$CDM and DDE. It is easy to see that a more negative value of $\omega$ in $\omega$CDM gives a smaller $f\sigma_8(z)$. However, even if we take $\omega=-2$ lying outside the $2\,\sigma$ confidence range, the value of $f\sigma_8(z)$ is still larger than the H22 measurement. This implies that just changing $\Lambda$CDM to $\omega$CDM used in Planck constraint can not clearly solve this small high-z tension. Interestingly, in the case of DDE, the value of $f\sigma_8(z)$ when we fix $\omega=-1$ and $\omega_a=-2$ at $1\,\sigma$ confidence level, which are allowed by the Planck constraint, can be well compatible with the H22 measurement. Subsequently, when we vary slightly $\omega$ from $-1$ to $-1.1$ and $-1.2$ in the DDE model, the value of $f\sigma_8(z)$ just has a very small global decrease. Furthermore, when we fix $\omega=-1$ and vary $\omega_a$ from -2 to -3 and -4, the value of $f\sigma_8(z)$ display a small decrease, which is more consistent with the H22 large scale structure measurement at $1\,\sigma$ confidence level. It is interesting that when fixing $\omega$, the difference of $f\sigma_8(z)$ values between $\omega_a=-3$ and -4 is much smaller than that between $\omega_a=-2$ and -3. These analytic results based on the Planck constraint reveal a fact that DDE can well solve the growth tension between Planck-2018 observations and the measurement derived from CMB lensing and galaxy clustering data at $z\sim4$.          

\section{Discussions and conclusions}
Recently, the CMB lensing signal made by 1.5 million galaxies is excitingly identified at $z\sim4$. However, under the assumption of $\Lambda$CDM, a small growth tension between the Planck-2018 observation and the combined observation of CMB lensing and galaxy clustering is seen over the $1\,\sigma$ confidence level. After exploring the parameter space of the DDE model constrained by the Planck-2018 data, we demonstrate that this discrepancy can be well solved by the DDE model within the $1\,\sigma$ confidence level. This means that this anomalous $f\sigma_8$ measurement at high redshift may give an evidence of evolution of DE. However, we find that $\omega$CDM can not solve this tension clearly.

It is worth noting that, to solve this anomaly, we only use the Planck data to constrain DDE but not use the H22 large scale structure observations. This implies that DDE will actually give a better consistency between two probes at the $1\,\sigma$ confidence level, if we also constrain DDE with the H22 data.        

This small high-z growth tension can invoke a discussion of alternative cosmological models. At the same time, we expect that future high-z large scale structure measurement can help confirm whether this growth anomaly really exists.

\section{Acknowledgments}
DW is supported by the National Nature Science Foundation of China under Grants No.11988101 and No.11851301.

\end{document}